\documentclass[superscriptaddress,showkeys,twocolumn,nofootinbib,longbibliography]{revtex4-1}

\usepackage{amsmath}
\usepackage{amssymb}
\usepackage{graphicx}
\usepackage{epsf}
\usepackage{slashed}
\usepackage{enumitem}
\usepackage[czech,english]{babel}
\usepackage[cp1250]{inputenc}
\usepackage{hyperref}
\usepackage{xcolor}

\usepackage[only,llbracket,rrbracket,llparenthesis,rrparenthesis]{stmaryrd} 
\usepackage{accsupp} 

\newcommand{\I}{\ensuremath{\mathrm{i}}}
\newcommand{\e}{\ensuremath{\mathrm{e}}}
\newcommand{\eL}{\mathcal{L}}
\renewcommand{\d}{\ensuremath{\mathrm{d}}}
\newcommand{\threevector}[1]{\boldsymbol{#1}}
\newcommand{\qm}[1]{``#1''} 

\usepackage{bbm} 

\usepackage[nodayofweek]{datetime}
\newdateformat{mydate}{\twodigit{\THEDAY}{ }\shortmonthname[\THEMONTH], \THEYEAR}
\usepackage[normalem]{ulem}

\begin{document}

\title{All finite-mass Dirac monopoles}

\newcommand{\affiliacePraha}{Institute of Experimental and Applied Physics, \\ Czech Technical University in Prague, Husova~240/5, 110~00 Prague~1, Czech Republic}

\newcommand{\affiliaceOpava}{Research Centre for Theoretical Physics and Astrophysics, Institute of Physics, \\ Silesian University in Opava, Bezru\v{c}ovo n\'{a}m\v{e}st\'{\i}~1150/13, 746~01 Opava, Czech Republic}

\author{Filip Blaschke}
\email{filip.blaschke@fpf.slu.cz}
\affiliation{Research Centre for Theoretical Physics and Astrophysics, Institute of Physics, \\ Silesian University in Opava, Bezru\v{c}ovo n\'{a}m\v{e}st\'{\i}~1150/13, 746~01 Opava, Czech Republic}
\affiliation{Institute of Experimental and Applied Physics, \\ Czech Technical University in Prague, Husova~240/5, 110~00 Prague~1, Czech Republic}

\author{Petr Bene\v{s}}
\email{petr.benes@utef.cvut.cz}
\affiliation{Institute of Experimental and Applied Physics, \\ Czech Technical University in Prague, Husova~240/5, 110~00 Prague~1, Czech Republic}

\begin{abstract}
We present a \qm{primitive} way of realizing finite-mass Dirac monopoles in $U(1)$ gauge theories involving a single non-minimally interacting scalar field. Typically, the energy density of this type of monopole is not concentrated at its core, but it is distributed in a spherical shell, as we illustrate on several exact solutions in the Bogomol'nyi--Prasad--Sommerfield (BPS) limit. We show that our construction can be interpreted as a limit of infinitely massive $W$ bosons coupled to electromagnetic field-strength via a dipole moment. Combining our approach with ideas of Weinberg and Lee, we present a general landscape of $U(1)$ gauge models that support a finite-mass Dirac monopole. In fact, all classical monopoles, i.e., Wu--Yang, 't~Hooft--Polyakov, Cho--Maison, etc., are special points on this landscape. 

\end{abstract}


\maketitle

\section{Bare monopoles \& Dressed monopoles}

In the classical Maxwell theory ($e$ denotes the electric charge)
\begin{equation}
\label{eq:maxwell}
\eL = -\frac{1}{2e^2}F_{\mu\nu}F^{\mu\nu}
\,,
\end{equation}
a magnetic monopole is a singular configuration of $U(1)$ gauge fields, i.e., the solution of $\vec \nabla \times \vec A = q\,\vec r /r^3$, with $q$ being a dimensionless constant. As is well known, the vector potential can be described everywhere except on a line going from the monopole to infinity, the so-called Dirac string  \cite{Dirac:1931kp}. For instance, a static monopole at the origin with a Dirac string lying on negative $z$-axis is given as
\begin{equation}
\label{eq:diracgauge}
A_i^{\rm D} = q \frac{\varepsilon_{i3j}x_j}{r(r+z)}
\,.
\end{equation}
The shape of the string can be changed via gauge transformation. 

The above description, which we dub a \emph{bare monopole}, is singular in three logically distinct ways. Namely, 1) there is an unphysical singularity along the Dirac string, 2) there is a physical singularity at the origin in both gauge fields and in the magnetic field $\vec B  = q\, \vec r /r^3 $ (we define $B_i \equiv \frac{1}{2} \varepsilon_{ijk} F_{jk}$) and 3) the energy density $\mathcal{E} = q^2 /(2 e^2 r^4)$ diverges at the origin in such a way that the total energy (i.e., the classical mass of the monopole) is infinite. 

These singularities are invariably clues that point to the incompleteness of the theoretical model. However, in so far as physics is concerned, they are not equivalently serious. The presence of Dirac string is simply a failure of global description of the $U(1)$ fibre bundle. The singularity in $\vec B$ is a result of taking the bird's eye point of view and describing the monopole as a point particle. By itself, this is not a problem as we can reasonably expect that in a more fundamental theory this singularity will be smoothed out by monopole's microscopic degrees of freedom. However, the real red flag is the absence of finite energy solutions. 

All in all, we are compelled to depart from the pure Maxwell theory. Fortunately, unlike the case of \emph{electric} monopoles, we do not have to abandon classical field theory to establish the existence of \emph{finite-mass Dirac monopoles}.

Let us now follow the (somewhat chronological) path that leads to finite-mass field-theoretical descriptions. 
First, there was an observation of Wu and Yang \cite{Wu:1976ge} that one can get rid of Dirac string by embedding \eqref{eq:diracgauge} into an $SU(2)$ gauge field ($q = 1$):
\begin{equation}
A_i^{\rm WY} = \frac{\varepsilon_{iaj}\sigma_a x_j}{2r^2} = A_i^{\rm D}U \sigma_3 U^{\dagger} - \I\, U \partial_i U^\dagger
\,,
\end{equation}
where $\sigma_a$ are the Pauli matrices and where
\begin{equation}
U = 
\left(\begin{array}{ll} 
           \cos\big(\theta/2\big) &
         - \sin\big(\theta/2\big)\,\e^{-\I\varphi} \\[2pt]
           \sin\big(\theta/2\big)\,\e^{ \I\varphi} &
\phantom{-}\cos\big(\theta/2\big)
\end{array}\right)
\end{equation}
is a (singular) $SU(2)$ gauge transformation that relates the Wu--Yang monopole $A_i^{\rm WY}$ and the Dirac monopole $A_i^{\rm D}$. By itself, however, this enhancement of symmetry is not sufficient to get rid of other singularities, neither in the $SU(2)$ \qm{magnetic} fields, nor in the energy density.

But, as was realized independently by 't~Hooft \cite{tHooft:1974kcl} and Polyakov \cite{Polyakov:1974ek}, the Wu--Yang monopole can be made completely regular via spontaneous symmetry breaking instigated by adjoint scalars. Loosely speaking, these additional fields condense at monopole's core and smooth out the $r = 0$ singularity -- the monopole becomes \emph{dressed}. 
Due to this field-dressing, the 't~Hooft--Polyakov monopole has none of the three singularities and represents the best classical description of a magnetic monopole.

However, as E.~Weinberg and K.~Lee pointed out \cite{Lee:1994sk} -- and what we want to propagate in this paper as well -- it is also possible to come up with finite-mass monopoles without resorting to non-Abelian gauge fields \eqref{eq:diracgauge}.

The key idea of \cite{Lee:1994sk} is to couple an $U(1)$ gauge field to a complex vector field $W_\mu$ with non-zero dipole moment tensor $d_{\mu\nu} = \I (W_\mu^\dag W_\nu - W_\mu W_\nu^\dag)$ that screens the bare monopole's charge via dipole moment interactions.
In other words:
\begin{eqnarray}
\eL &=&
- \frac{1}{4g^2} F_{\mu\nu}^2
- \frac{\eta}{2} d_{\mu\nu} F^{\mu\nu}
- \frac{\chi g^2}{4}d_{\mu\nu}^2
\nonumber \\ && {}
- \frac{1}{2} \big|D_\mu W_\nu-D_\nu W_\mu\big|^2
+ m^2 \big|W_\mu\big|^2
\,,
\end{eqnarray}
where $D_\mu W_\nu = (\partial_\mu+\I A_\mu) W_\nu$. The parameter $\eta$ measures the strength of the dipole interactions with external el.~mag.~field, while $\chi$ measures the dipole-dipole self-interactions. For a static configuration, the energy density can be rewritten as
\begin{eqnarray}
\mathcal{E} &=&
\frac{\chi}{4}\bigg(g\, d_{ij} + \frac{\eta}{\chi\, g}F_{ij}\bigg)^2
+ \frac{1}{4g^2}\bigg(1- \frac{\eta^2}{\chi}\bigg)F_{ij}^2
\nonumber \\ && {}
+ \frac{1}{2}\big|D_iW_j-D_j W_i\big|^2
+ m^2 \big|W_i\big|^2
\,.
\end{eqnarray}
In order to cancel the singularity in $F_{ij}$ we need $d_{ij}$ to have the same type of singularity and -- as we see -- we must require $\eta^2 = \chi$. However, by itself this is not enough to make $\mathcal{E}$ regular  \cite{Lee:1994sk}. Critically, the mass term would still give us $\sim 1/r^2$. Furthermore, simply setting $m = 0$ from the beginning would not help either. This can be seen in the limit $\eta = \lambda = 1$, where the above model can be rewritten as pure $SU(2)$ model upon identifications $A_\mu = A_\mu^3$ and $W_\mu = (A_\mu^1 +\I A_\mu^2) / (\sqrt{2} g)$. The static solution is gauge-equivalent to Wu--Yang monopole which we know has divergent energy.

Thus a further field must be added, namely a real scalar field $\phi$ that develops a non-zero expectation value at the vacuum, but which vanishes at the monopole's core. The idea is to make mass of $W_\mu$ dependent on $\phi$ in such a way that $m^2(\phi) \to m^2(0) = 0$ as $r \to 0$.

All in all, the Weinberg--Lee's procedure for regularising magnetic monopole can be encompassed by the following Lagrangian:
\begin{eqnarray}
\label{eq:weinberg}
\eL_{\rm WL} &=&
- \frac{1}{4g^2} F_{\mu\nu}^2
- \frac{\eta}{2} d_{\mu\nu} F^{\mu\nu}
- \frac{\chi g^2}{4}d_{\mu\nu}^2
\nonumber \\ && {}
- \frac{1}{2} \big|D_\mu W_\nu-D_\nu W_\mu\big|^2
+ m^2(\phi) \big|W_\mu\big|^2
\nonumber \\ && {}
+ \frac{1}{2}(\partial_\mu \phi)^2
- \frac{\lambda}{2} \big(\phi^2-v^2\big)^2
\,.
\end{eqnarray}
The total energy of a monopole solution is finite and regular if we take $\eta^2 = \chi$ and make an appropriate choice of $m^2(\phi)$. In \cite{Lee:1994sk}, these solutions were dubbed \emph{non-topological magnetic monopoles}, as the mechanism that renders monopoles in these models to have finite classical masses has nothing to do with topological reasoning \'a la 't~Hooft--Polyakov. Nevertheless, the 't~Hooft--Polyakov monopole is, in fact, a special point in the above family of models, namely $\eta = \chi = 1$ and $m^2(\phi) = g^2 \phi^2$. Indeed, with these assignments \eqref{eq:weinberg} becomes a unitary gauge of $SU(2)$ theory with an adjoint scalar $\Phi = \frac{1}{2} \sigma_3 \phi$.

\begin{figure}
\begin{center}
\includegraphics[width=1\columnwidth,trim = 45 0 30 0]{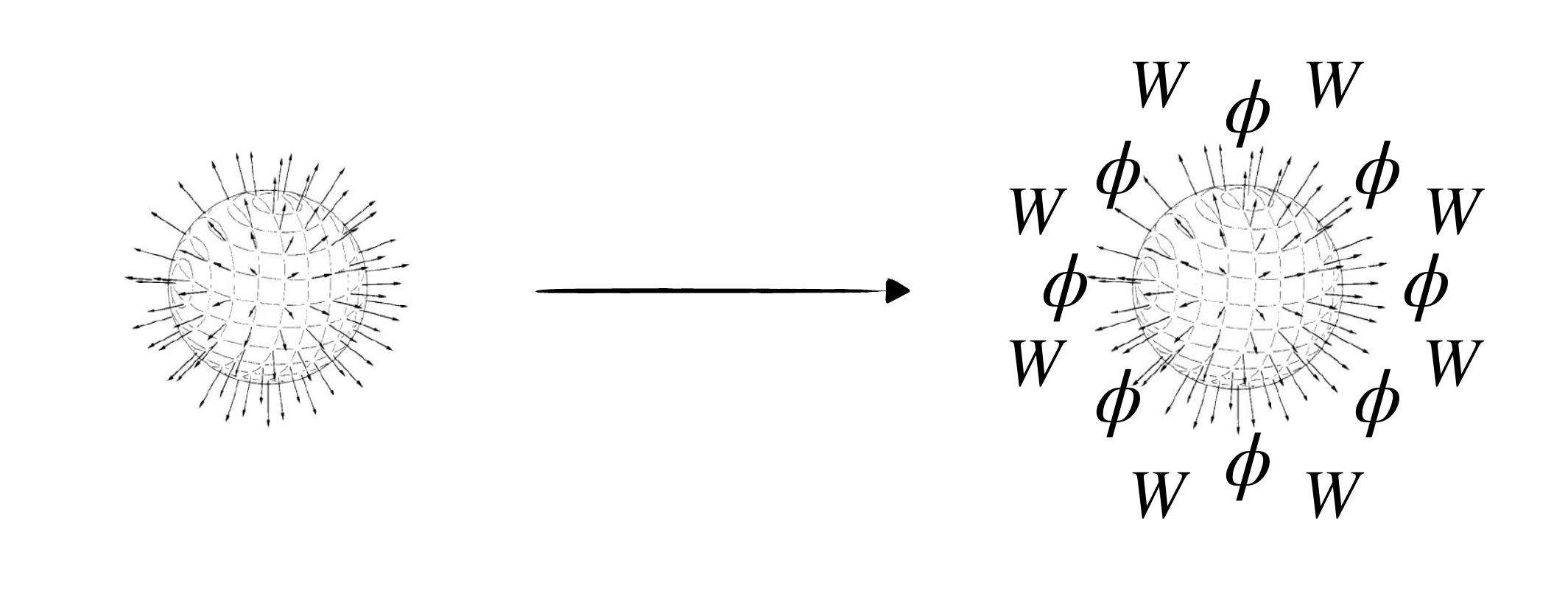}
\caption{Transition from a bare monopole of \eqref{eq:maxwell} to a \qm{dressed} monopole of \eqref{eq:weinberg}.}
\label{fig:weinberg}
\end{center}
\end{figure}

To summarise, Eq.~\eqref{eq:weinberg} gives us finite-mass magnetic monopoles that, compared to bare monopoles of Eq.~\eqref{eq:maxwell}, are dressed in complex vector fields $W_\mu$ with explicit dipole moment interactions and a scalar field $\phi$ that controls the mass of $W_\mu$'s, making it effectively zero at the monopole's core. This idea can be depicted as in Fig.~\ref{fig:weinberg}.

In the next section, we introduce a \qm{primitive} way of regularizing magnetic monopole's energy that is seemingly unrelated to these arguments. Our construction will lead us to novel solutions that can be described exactly in the BPS limit $\lambda \to 0$. However, in section \ref{sec:3} we will unify our and Weinberg--Lee's points of view into a much broader landscape of theories that support finite-mass Dirac monopoles.

\section{A primitive way}

The core idea of this paper is to have a real scalar field $\phi$ interact directly with the el.~mag.~field via non-canonical kinetic term:
\begin{equation}
\label{eq:primitive}
\eL =
- \frac{h^{\prime\,2}(\phi/v)}{4g^2} F_{\mu\nu}^{2}
+ \frac{1}{2} (\partial_\mu\phi)^2
- \frac{\lambda}{2}\big(\phi^2-v^2\big)^2
\,.
\end{equation} 
From the outset, there seems to be no need for complex vector field $W_\mu$ and dipole moment interactions (but see Sec.~\ref{sec:3}). Indeed, this model can yield finite-energy magnetic monopole solutions for appropriate choices of so-far arbitrary function $h^\prime(\phi/v)$ (this notation with the prime denoting a derivative becomes obvious when we discuss the BPS limit). Without loss of generality we take the normalization $h^\prime(1) = 1$.

\begin{figure}
\begin{center}
\includegraphics[width=1\columnwidth,trim = 35 0 35 0]{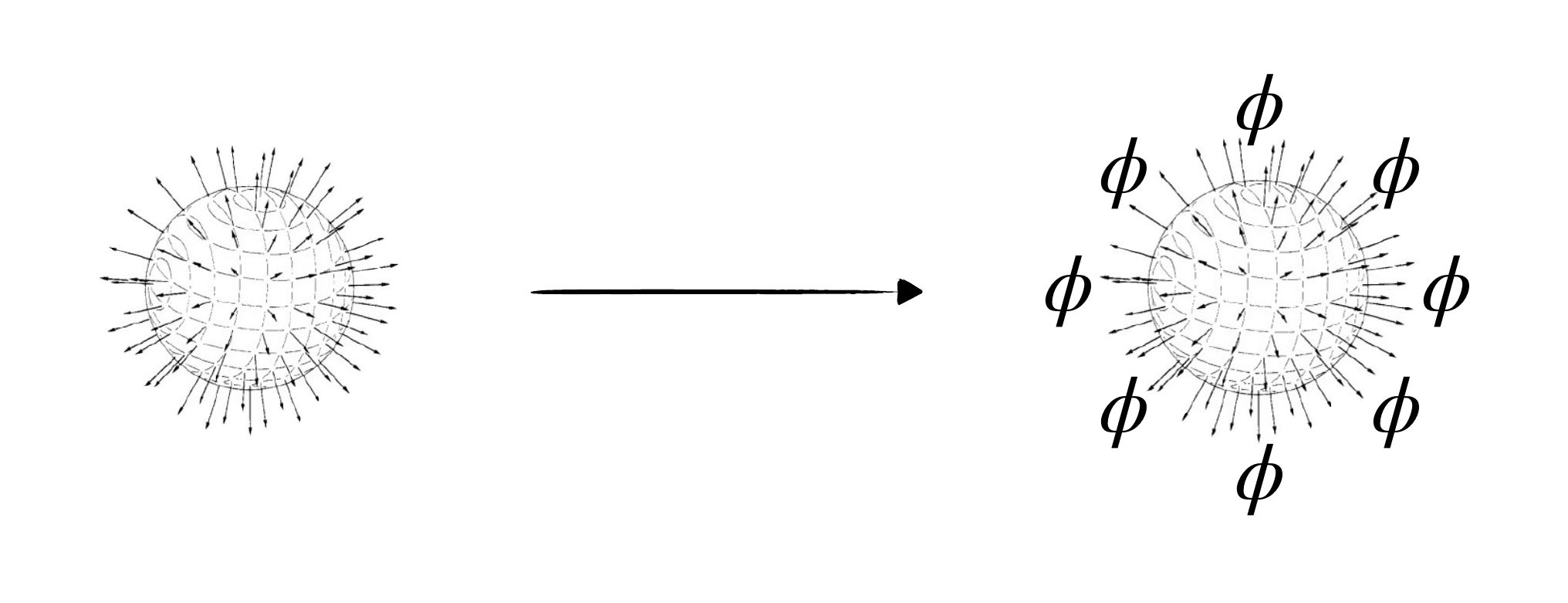}
\caption{Transition from a bare monopole of \eqref{eq:maxwell} to a \qm{dressed} monopole of \eqref{eq:primitive}.}
\label{fig:primitive}
\end{center}
\end{figure}

We can depict this primitive way of a field-dressing by means of just a scalar field as in Fig.~\ref{fig:primitive}.

Let us analyze the full equations of motion for a static configuration,
\begin{subequations}
\begin{gather}
\vec\nabla \times \Big(h^{\prime\,2}(\phi/v) \, \vec B\Big) = 0
\,,
\\
\vec \nabla^2 \phi = \frac{h^\prime(\phi/v) \, h^{\prime\prime}(\phi/v)}{g^2 v} \vec B^2 +2 \lambda \phi \big(\phi^2-v^2\big)
\,.
\end{gather}
\end{subequations}
We see that in the monopole background, $\vec B = q \, \vec r/r^3$, the first equation is solved identically if the scalar field depend only on the (dimensionless) radius, i.e., $\phi \equiv v \sigma (\rho)$, where $\rho = g v r$. The second equation (subject to the boundary conditions $\sigma(0) = 0$ and $\sigma(\infty) = 1$) then reads
\begin{equation}
\frac{1}{\rho^2}\partial_{\rho}\big(\rho^2 \partial_\rho \sigma\big) = 
q^2\frac{h^\prime(\sigma) \, h^{\prime\prime}(\sigma)}{\rho^4}
+ \frac{2\lambda}{g^2} \sigma \big(\sigma^2-1\big)
\,.
\end{equation}

From the expression for the energy density
\begin{equation}
\frac{\mathcal{E}}{g^2 v^4} = 
q^2\frac{h^{\prime\,2}(\sigma)}{2\rho^4}
+ \frac{1}{2}(\partial_\rho \sigma)^2
+ \frac{\lambda}{2g^2}\big(\sigma^2-1\big)^2
\end{equation}
it is clear that a necessary condition for $\mathcal{E}$ to be regular at origin is $h^\prime(0) = 0$. It is not a sufficient condition, though, as can be seen by assuming a leading power dependence $h^\prime \sim h_0 \sigma^\alpha$ as $\sigma \to 0$. In this case one must additionally assume $\alpha \in [1,2]$ in order to maintain regularity of $\mathcal{E}$. The asymptotic behaviour of $\sigma$ and $\mathcal{E}$ near the origin then falls into two categories:
\begin{subequations}
\begin{eqnarray}
\label{eq:asympt1}
1<\alpha \leq 2 &\ :&\hspace{6mm}
\left\{\begin{array}{ccl}
\sigma &\sim&
\displaystyle
\bigg[\frac{\rho}{q h_0(\alpha-1)}\bigg]^{\frac{1}{\alpha-1}}
\,,
\\[10pt]
\displaystyle
\frac{{\mathcal E}}{g^2v^4} &\sim&
\displaystyle
\bigg[ \frac{\rho^{2-\alpha}}{q h_0(\alpha-1)^{\alpha}} \bigg]^{\frac{2}{\alpha-1}}
\,,
\end{array}\right.
\hspace{10mm}
\end{eqnarray}
and
\begin{eqnarray}
\label{eq:asympt2}
\alpha = 1 &\ :&\hspace{6mm}
\left\{\begin{array}{ccl}
\sigma &\sim&
\displaystyle
\sigma_0
\exp\!\bigg(-\frac{|q h_0|}{\rho}\bigg)
\,,
\\[10pt]
\displaystyle
\frac{\mathcal{E}}{g^2v^4} &\sim&
\displaystyle
\frac{q^2 h_0^2\sigma_0^2}{\rho^4}
\exp\!\bigg(-2\frac{|q h_0|}{\rho}\bigg)
\,.
\end{array}\right.
\hspace{8mm}
\end{eqnarray}
\end{subequations}
In the first category ($1< \alpha \leq 2$) both $\sigma$ and $\mathcal{E}$  fall off in the origin as a power law with a positive exponent. In the second category ($\alpha = 1$), which is just a limit case $\alpha \to 1$, both $\sigma$ and $\mathcal{E}$ develop non-analytic profiles $\sim \e^{-1/\rho}$. As we shall see, this behaviour is indicative of a maximal region around monopole's core where almost no energy is stored. Thus, this kind of monopole is \qm{hollow} (a term which we adopt from \cite{Bazeia:2018fhg}). It should be also pointed out, that only in the case $\alpha=2$, the energy density does not vanish at the origin. In all other cases, the energy is stored in a spherical shell.

\subsection{The BPS limit}

In the BPS limit $\lambda \to 0$, the energy density can be completed into a total square as
\begin{equation}
\mathcal{E} =
\frac{1}{2}\bigg(\vec \nabla \phi - \frac{h^\prime(\phi/v)}{g}\vec B\bigg)^2
+ \frac{v}{g} \vec\nabla \big(h(\phi/v)\,\vec B\big)
\,,
\end{equation}
where we used $h(\phi/v) \, \vec \nabla \cdot \vec B = h(0) q \delta^3(\vec r) = 0$. Thus, we demand that the primitive function of $h^\prime$ is zero at each zero of $\phi$, namely $h(0) = 0$. Notice that this fixes the constant of integration. With this additional condition, one can easily derive the formula for mass of the BPS monopole:
\begin{equation}
\label{eq:mass}
M_{\rm BPS} \ = \ \frac{4\pi v q}{g}  h(1) \ = \ \frac{4\pi v q}{g} \int_{0}^{1} \! h^\prime(\sigma)\,\d\sigma
\,. 
\end{equation}

\begin{figure}
\begin{center}
\includegraphics[width=1.03\columnwidth]{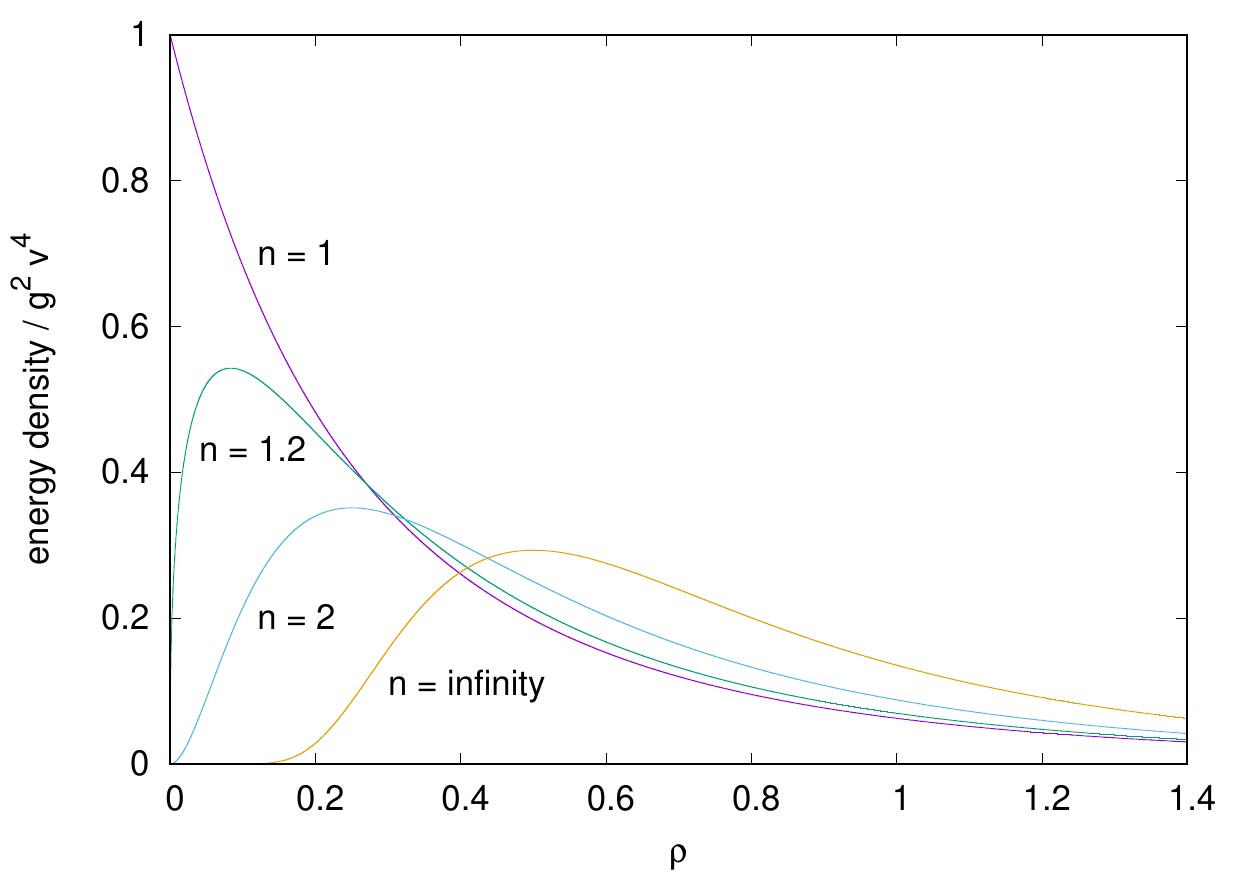}
\caption{Energy density profiles \eqref{solutions} for various $n$. We take $q = 1$.}
\label{fig:one}
\end{center}
\end{figure}

Let us illustrate the BPS solutions on a particular family of functions
\begin{equation}
h^\prime(\sigma) = \sigma^{1+1/n} \,,
\end{equation}
where $n \geq 1$. The BPS equation $\vec \nabla \phi = h^\prime(\phi/v)\,\vec B/g$ can be easily integrated. For instance, taking a single monopole at the origin,\footnote{Exact multi-monopole solutions can be readily obtained as well, simply by taking $\vec B = q \sum_i (\vec r - \vec r_i)/|\vec r - \vec r_i|^3$, where the sum goes over positions $\vec r_i$ of the monopoles.} i.e., $\vec B = q \, \vec r/r^3$, the result reads
\begin{subequations}
\label{solutions}
\begin{eqnarray}
\phi
= \frac{v}{(1 + \frac{q}{n \rho})^n}
\hspace{5mm}
&\mbox{and}&
\hspace{5mm}
\frac{\mathcal{E}}{g^2 v^4}
= \frac{q^2}{\rho^4(1+\frac{q}{n \rho})^{2n+2}}
\,,
\hspace{8mm}
\end{eqnarray}
in full accordance with our asymptotic formulas \eqref{eq:asympt1}. The limit $n \to \infty$ takes us directly to the \qm{hollow} monopole case:
\begin{eqnarray}
\phi = v\, \e^{-q/\rho}
\hspace{5mm}
&\mbox{and}&
\hspace{5mm}
\frac{\mathcal{E}}{g^2 v^4} = \frac{q^2}{\rho^4}\e^{-2q/\rho}
\,,
\end{eqnarray}
\end{subequations}
corresponding to the asymptotic behaviour \eqref{eq:asympt2}. We display the energy densities \eqref{solutions} in Fig.~\ref{fig:one}. The mass of the monopole follows from \eqref{eq:mass} as
\begin{equation}
M
= \frac{4\pi v q}{g} \, \frac{n}{2n+1}
\,.
\end{equation}

\section{Outlook: Broadening the landscape}
\label{sec:3}

At first, it seems that our primitive approach of the previous section runs orthogonal to the Weinberg--Lee's ideas of charge screening via dipole moment interactions. However, there is a natural connection between the two that results from the following observation. We can rewrite Eq.~\eqref{eq:primitive} as
\begin{eqnarray}
\label{eq:frozen}
\eL
&=&
- \frac{1}{4g^2} F_{\mu\nu}^2
- \frac{\eta(\phi/v)}{2} \tilde{d}_{\mu\nu} F^{\mu\nu}
- g^2 \frac{\chi(\phi/v)}{4} \tilde{d}_{\mu\nu}^2
\nonumber \\ && {}
+ \frac{1}{2} (\partial_\mu \phi)^2
- \frac{\lambda}{2} \big(\phi^2-v^2\big)^2
\,.
\end{eqnarray}
Indeed, upon eliminating $\tilde{d}_{\mu\nu}$ via its equation of motion $\partial_{\tilde{d}_{\mu\nu}}\eL = 0$, we obtain
\begin{equation}
\eL = 
- \frac{1}{4g^2} \bigg[1-\frac{\eta^2(\phi/v)}{\chi(\phi/v)}\bigg] F_{\mu\nu}^2
+ \frac{1}{2} (\partial_\mu \phi)^2
- \frac{\lambda}{2} \big(\phi^2-v^2\big)^2
\,.
\end{equation}
Hence we return to the original Lagrangian \eqref{eq:primitive} if\footnote{One can absorb either $\eta$ or $\chi$ into the definition of $\tilde{d}_{\mu\nu}$, hence there is no ambiguity in the correspondence between \eqref{eq:frozen} and \eqref{eq:primitive}.}
\begin{equation}
\label{eq:relation}
\eta^2 = \chi \big(1-h^{\prime\, 2}\big) \,.
\end{equation}
Furthermore, we can identify $\tilde{d}_{\mu\nu}$ with the dipole moment tensor of the vector field $W_\mu$ in the limit of infinite mass.\footnote{By this, we mean that $W_\mu$ fields are frozen so that their kinetic term vanishes. Moreover, a similar Lagrangian as in Eq.~\eqref{eq:frozen} can be obtained by starting with the Weinberg--Lee model \eqref{eq:weinberg} and dropping the kinetic term $|D_{[\mu}W_{\nu]}|^2$ there. Of course, the functions $m(\phi/v)$, $\chi(\phi/v)$ and $\eta(\phi/v)$ are different, but the structure is similar, which is what we emphasize. The mass term $m^2 |W|^2$, that is present in \eqref{eq:weinberg} but not in \eqref{eq:frozen}, can be accounted for (at least in local patches) by assigning $\tilde{d}_{\mu\nu} = d_{\mu\nu} + \frac{m(\phi/v)}{g^2\chi(\phi/v)} g_{\mu\nu} \sqrt{-W_\rho^\dagger W^\rho}$, where $d_{\mu\nu} = \I (W_\mu^\dagger W_\nu - W_\nu^\dagger W_\mu)$ as in \eqref{eq:weinberg}.}


It is in this sense that our primitive approach falls into the same purview as the Weinberg--Lee's procedure, namely that the finiteness of the monopole's mass is realized by screening of the magnetic charge via dipole moment interactions. 

There is, however, one important difference. Notice that our approach calls for field-dependent $\eta$ and $\chi$ couplings that were, however, kept constant in \eqref{eq:weinberg}. From the formula \eqref{eq:relation}, we see that at the monopole's position $\eta^2(0) = \chi(0)$, namely the same condition required by the Weinberg--Lee scheme for eliminating the singularity in energy density. On the other hand, in the vacuum $\eta(1) = 0$, i.e., the dipole moment interactions are shut down. This is physically different process then having field-dependent mass of $W_\mu$ fields.

Thus, we see that ours and Weinberg--Lee's approach are not equivalent, but they are a special cases of a yet more general procedure that can be subsumed by the following landscape of $U(1)$ gauge theories:
\begin{eqnarray}
\label{eq:landscape}
\eL &=&
- \frac{f_1^2(\phi/v)}{4g^2} F_{\mu\nu}^2
- \frac{\eta(\phi/v)}{2} d_{\mu\nu} F^{\mu\nu}
- \frac{\chi(\phi/v) g^2}{4} d_{\mu\nu}^2
\nonumber \\ && {}
- \frac{f_2^2(\phi/v)}{2} \big|D_\mu W_\nu-D_\nu W_\mu\big|^2
+ m^2(\phi/v) \big|W_\mu\big|^2
\nonumber \\ && {}
+ \frac{1}{2} (\partial_\mu \phi)^2
- \frac{\lambda}{2} \big(\phi^2-v^2\big)^2
\,.
\end{eqnarray}
In this landscape both approaches are combined: Both mass and dipole couplings are taken as functions of $\phi/v$ (again, some redundancy exists as one of those functions can be eliminated by redefinition of $W_\mu$ fields). Furthermore, the kinetic terms of both $A_\mu$ and $W_\mu$ have field-dependent pre-factors.

Eq.~\eqref{eq:landscape} represents a vast territory. It contains Wu--Yang, 't~Hooft--Polyakov, Weinberg--Lee's and our primitive monopoles as a special cases. As already noted in \cite{Lee:1994sk}, by adding additional real vector field $Z_\mu$ together with appropriate interaction terms, the above theory can be easily mapped to the bosonic sector of the Electroweak model (or some of its modification). In this way, both the so-called Cho--Maison monopole \cite{Cho:1996qd} and all its finite-mass incarnations \cite{Cho:2013vba, Ellis:2016glu, Blaschke:2017pym} are naturally contained in this conceptual scheme.

Further, if we allow for more scalar fields, we can incorporate various non-linear electrodynamics theories. For instance, to obtain a Born--Infeld electromagnetism, we simply use a non-dynamical scalar field, say $\varphi$, as a Lagrange multiplier:
\begin{eqnarray}
\eL &\supset&
- \frac{\varphi}{4g^2} F_{\mu\nu}^2 
- \frac{\beta^2}{2}\bigg(\varphi+\frac{1}{\varphi}\bigg)
+ 2\beta^2
\\
&\xrightarrow[\partial_\varphi \eL = 0]{}&
\beta^2 \Bigg(1-\sqrt{1+\frac{1}{2g^2 \beta^2}F_{\mu\nu}^2}\Bigg)
\,.
\end{eqnarray} 

Notice that \eqref{eq:landscape} also contains non-Abelian theories with non-minimally coupled adjoint scalars. This can be seen by writing down such models in unitary gauge and figuring our relations between field-dependent functions. In this way, the following sub-landscape can be derived:
\begin{eqnarray}
\label{eq:noncan}
\eL &=&
\frac{v^2}{2} f_1^2(\phi/v) \big(D_\mu \threevector{n}\big)^2
\nonumber \\ && {}
- \frac{1}{4g^2} \bigg[f_2^2(\phi/v) \threevector{F}_{\mu\nu}^2 
+ f_3(\phi/v) \big(\threevector{n} \cdot \threevector{F}_{\mu\nu}\big)^2\bigg]
\nonumber \\ && {}
+ \frac{1}{2} (\partial_\mu \phi)^2
- \frac{\lambda}{2} \big(\phi^2-v^2\big)^2
\,,
\hspace{10mm}
\end{eqnarray}
where we decomposed the adjoint triplet $\threevector{\phi} = \phi \, \threevector{n}$ into a scalar field $\phi$ and a unit iso-vector $\threevector{n}^2 = 1$. Here, we employ $SO(3)$ notation.

We plan to investigate the monopole solutions of these models in a separate publication.
 



\acknowledgments

F.~B.~would like to thank M.~Eto and M.~Arai for valuable discussions and express gratitude for the institutional support of the Research Centre for Theoretical Physics and Astrophysics, Institute of Physics, Silesian University in Opava and to the Institute of Experimental and Applied Physics, Czech Technical University in Prague. P.~B.~is grateful to FSM and A\v{s}tar \v{S}eran for support. This work was supported by the grant of the Ministry of Education, Youth and Sports of the Czech Republic no.~LTT17018.


\providecommand{\href}[2]{#2}\begingroup\raggedright\endgroup

\end{document}